\RequirePackage{ifpdf}
\documentclass{pCHEF}

\title{Pandora Particle Flow Algorithm}

\ShortTitle{Pandora Particle Flow Algorithm}

\author{\speaker{J. S. Marshall}\\
        Cavendish Laboratory, University of Cambridge, J. J. Thomson Avenue, Cambridge, UK.\\
        E-mail: \email{marshall@hep.phy.cam.ac.uk}}
\author{M. A. Thomson\\
        Cavendish Laboratory, University of Cambridge, J. J. Thomson Avenue, Cambridge, UK.\\
        E-mail: \email{thomson@hep.phy.cam.ac.uk}}

\abstract{A high-energy e$^{+}$e$^{-}$ collider, such as the ILC or CLIC, is arguably the best option to complement and extend the LHC physics programme. A lepton collider will 
allow for exploration of Standard Model Physics, such as precise measurements of the Higgs, top and gauge sectors, in addition to enabling a multitude of New Physics searches. 
However, physics analyses at such a collider will place unprecedented demands on calorimetry, with a required jet energy resolution of \mbox{$\sigma_{E}/E\lesssim3.5\,\%$.} To meet
these requirements will need a new approach to calorimetry.

The particle flow approach to calorimetry requires both fine granularity detectors and sophisticated software algorithms. It promises to deliver unparalleled jet energy resolution 
by fully reconstructing the paths of individual particles through the detector. The energies of charged particles can then be extracted from precise inner detector tracker
measurements, whilst photon energies will be measured in the ECAL, and only neutral hadron energies (10\% of jet energies) will be measured in the HCAL, largely avoiding the 
typically poor HCAL resolution.

This document introduces the Pandora particle flow algorithms, which offer the current state of the art in particle flow calorimetry for the ILC and CLIC. The performance of the
algorithms is investigated by examining the reconstructed jet energy resolution and the ability to separate the hadronic decays of $W$ and $Z$ bosons.}

\FullConference{Calorimetry for High Energy Frontiers - CHEF 2013,\\
		April 22-25, 2013\\
		Paris, France}

\usepackage{CHEF2013_definitions}

\begin{document}

\section{Particle Flow Calorimetry}
At a future high-energy lepton collider, such as the International Linear Collider (ILC)~\cite{ILC} or Compact Linear Collider (CLIC)~\cite{CLIC_PhysDet_CDR, CLICCDR_vol3}, many 
interesting physics processes will produce final states that consist of multiple jets, often accompanied by charged leptons and/or missing transverse momentum. The ability to 
accurately reconstruct the invariant masses of the jets proves vital in order to perform precision physics measurements: the masses are needed for both reconstruction and 
identification of events. The jet energy resolution goal at the ILC or CLIC is that it 
should allow separation of the hadronic decays of $W$ and $Z$ bosons via the reconstruction of the di-jet invariant masses. This sets a challenging jet energy resolution target 
of \mbox{$\sigma_{E}/E\lesssim3.5\,\%$} for $50-500$\,GeV jets at the ILC and for up to 1.5\,TeV jets at CLIC. This goal is unlikely to be achieved using a traditional 
approach to calorimetry~\cite{Thomson:2009rp}.

Measurements of jet fragmentation at LEP provide detailed information about the particle composition of jets~\cite{Knowles:1997dk,LEPExpt2}. In a typical jet, approximately 62\,\% 
of the energy is carried by charged particles (mainly hadrons), whilst 27\,\% is carried by photons, 10\,\% by long-lived neutral hadrons and 1.5\,\% by neutrinos. A traditional 
approach to calorimetry would measure the jet energy via the energies deposited in the electromagnetic and hadronic calorimeters (ECAL and HCAL). For a typical jet, this means 
that 72\,\% of the energy would be measured in the HCAL, with a typical resolution of \mbox{$\gtrsim 55\,\%/\sqrt{E/\mathrm{GeV}}$}, greatly limiting the achievable jet energy 
resolution.

\begin{figure}[h!] 
\begin{center}
\includegraphics[width=.82\textwidth]{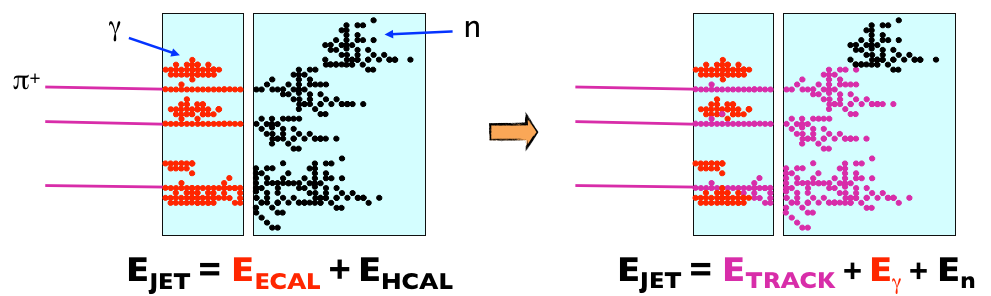}
\caption{The transition from traditional calorimetry to fine granularity particle flow calorimetry.}
\label{fig1} 
\end{center}
\end{figure}

The particle flow approach to calorimetry aims to improve the jet energy resolution by tracing the paths of individual particles through the detector, collecting together the
energy deposits left in each subdetector system, as illustrated in Figure~\ref{fig1}. The energy and momentum for each particle can then be extracted from the subdetector system 
in which we expect the measurement to be most accurate. Charged particle momenta can be measured precisely in the inner detector tracker, whilst photon energies can be obtained 
from the energy deposits in the ECAL, with typical resolution \mbox{$\lesssim 20\,\%/\sqrt{E/\mathrm{GeV}}$}. The HCAL is then only used to measure the 10\,\% of the jet energy 
carried by long-lived neutral hadrons. Particle flow calorimetry can therefore offer a significant improvement to jet energy measurements, but it relies on accurate pattern recognition
techniques to collect together the energy deposits from individual particles.

\section{Realising Particle Flow Calorimetry}
\begin{figure}[h!] 
\begin{center}
\includegraphics[width=.9\textwidth]{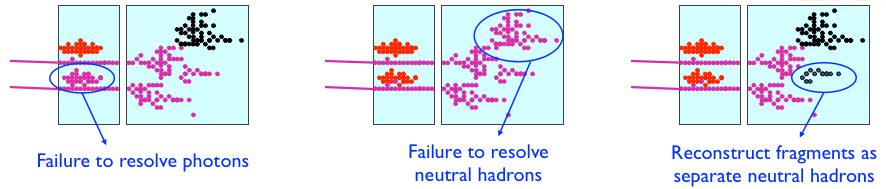}
\caption{Possible sources of confusion in a fine granularity particle flow reconstruction.}
\label{fig2} 
\end{center}
\end{figure}

Particle flow calorimetry requires the energy depositions from individual particles to be traced through the detector and cleanly separated from the depositions of other particles.
This reconstruction of the individual particles in the event requires both fine granularity calorimeters and sophisticated software algorithms. Failure to provide sufficient
sampling points in the calorimeter, or a lack of sophistication in the pattern recognition algorithms, will likely lead to ${confusion}$ in the particle reconstruction. 
Figure~\ref{fig2} illustrates the different possible sources of confusion in identifying the particles, which will lead to degradation of the jet energy measurement. It
is this confusion, rather than the intrisic calorimetric resolution, which proves to be the limiting factor for particle flow calorimetry:

\begin{itemize}
\item{Failure to resolve neutral particles (photons or neutral hadrons) from nearby charged hadrons will result in loss of energy. The energy deposits of the neutral particle will
be added to those of the charged particle, but the charged particle four-vector will be reconstructed using measurements from the inner detector tracker.}
\item{Failure to associate all the calorimeter energy deposits from a charged particle with the correct inner detector track will lead to double counting of energy. The unassociated
calorimeter energy deposits will be used to create a fake neutral particle, whilst the track will still be used to provide the four-vector for the true charged particle.}
\end{itemize}

In order to fully exploit particle flow calorimetry, the confusion must be reduced to the lowest possible level. This places constraints on both the calorimeter hardware and the
software pattern recognition algorithms. In terms of the hardware, accurate inner detector tracking is vital, alongside calorimeters that can longitudinally separate electromagnetic
and hadronic showers. The ECAL must therefore have a large ratio of radiation length to nuclear interaction length. Its Moli\`{e}re radius must also be small, in order to reduce the
transverse spread of electromagnetic showers and aid the separation of photons from nearby charged hadrons. The transverse and longitudinal sampling in the ECAL must be sufficient
to allow separate clustering and identification of electromagnetic showers by the particle flow algorithms.

\begin{figure}[h!] 
\begin{center}
\includegraphics[width=.4\textwidth]{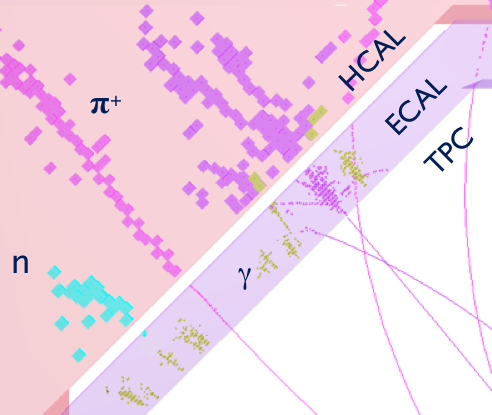}
\caption{A section of a typical 250\,GeV jet in ILD. The subdetectors and several particles are labelled.} 
\label{fig3} 
\end{center}
\end{figure}

The HCAL must offer longitudinal and transverse segmentation, sufficient to allow separation of neutral hadrons from nearby charged particles. The HCAL should also aim to fully contain
hadronic showers, so a small nuclear interaction length is desirable. It will be a rather large component of the detector, so its cost and structural properties are also
of importance. One detector concept that was designed with particle flow calorimetry in mind is ILD~\cite{ILD}. Figure~\ref{fig3} shows a section of a typical
250\,GeV jet in ILD, with labels identifying a number of the constituent particles. The Figure shows inner detector tracks, representing the paths of charged particles in the
Time Projection Chamber (TPC). These tracks can be associated by eye to calorimeter energy deposits in the ECAL and HCAL. Also visible by eye are photon energy deposits in the
ECAL and a neutron in the HCAL. The challenge then is to develop software algorithms that can match the human eye for the quality of its particle reconstruction.

Fine granularity particle flow calorimetry lives or dies on the quality of the reconstruction of its particles. High performance software is required, in terms of algorithm
sophistication, with reliable implementation, and in terms of the CPU and memory usage (they must process complicated events with many hits). The algorithms must be able to exploit
the granularity of the calorimeters, whilst making very few mistakes and processing events quickly. The most sensible approach is to implement a large number of `decoupled'
pattern-recognition algorithms, each of which looks to reconstruct specific particle topologies, whilst carefully avoiding causing confusion.

\begin{figure}[h!] 
\begin{center}
\includegraphics[width=.82\textwidth]{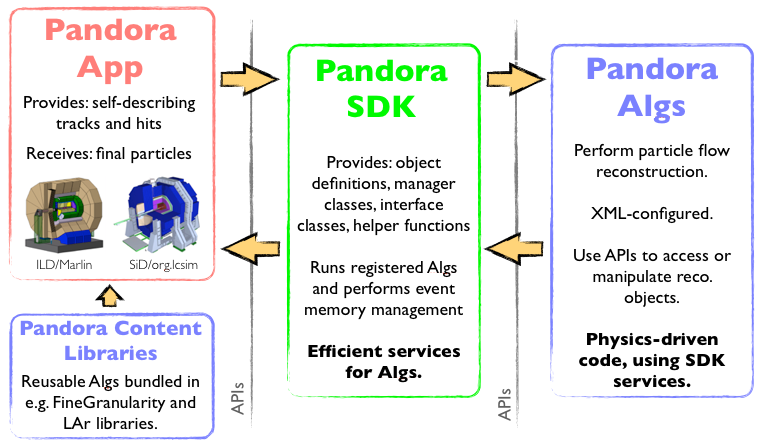}
\caption{The division of the Pandora reconstruction software into client application, central framework and algorithms. The client application provides the input to
the reconstruction and receives the final particles. The framework handles the memory management, whilst the algorithms implement the pattern recognition.} 
\label{fig4} 
\end{center}
\end{figure}

The need to implement a large number of efficient algorithms motivates the need for a central sofware framework, which can take care of memory-management and book-keeping issues,
allowing each algorithm to remain simple and focussed on its particular pattern-recognition task. Almost all ILC/CLIC studies use code developed with the Pandora C++ Software
Development Kit (SDK)~\cite{Marshall:2012hh}. The PandoraSDK is a robust and efficient framework for developing and running pattern-recognition algorithms for particle flow reconstruction.
It consists of a library and a number of carefully designed Application Programming Interfaces (APIs). Using the PandoraSDK means that the reconstruction is cleanly divided into
three sections, which communicate via the APIs. This is illustrated in Figure~\ref{fig4}.

A Pandora client application uses the APIs to pass details of the tracks and calorimeter cells in an event to Pandora, which then creates and manages named lists of its own
self-describing reconstruction objects. These objects can be accessed by the Pandora algorithms, which perform the particle flow reconstruction. The important point is that
the algorithms can only access and/or manipulate the reconstruction objects by requesting services from Pandora, e.g. there are APIs for creating clusters, or merging multiple
clusters together. As a result of this software-engineering approach, with controlled access to the reconstruction objects, Pandora can perform the memory-management and the 
algorithms become more efficient, with a `cleaner' implementation.

\section{Pandora Particle Flow Algorithms}
The reconstruction of events in a fine granularity detector, such as ILD, uses over 60 different Pandora algorithms. These algorithms are well-understood and have been documented
in \cite{Thomson:2009rp,Marshall:2012ry}. The basic reconstruction operations performed by the default set of Pandora algorithms is briefly summarised below:

\begin{itemize}
\item{Calorimeter cells are clustered using a simple cone-based clustering algorithm, working outwards in the calorimeters from the front of the ECAL to the back of the HCAL.
Clusters can be seeded by the projection of inner detector tracks to the front face of the ECAL.}
\item{The clustering algorithm is configured so that it tends to split up the energy deposits from individual particles, rather than risk accidentally merging particles so
early in the reconstruction. The resulting proto-clusters are then carefully merged together by a series of algorithms that implement well-motivated topological rules. The
fine granularity and tracking capabilities of the calorimeter are exploited to merge clusters whilst making very few mistakes.}
\item{The calorimeter clusters are carefully associated to the inner detector tracks, by comparing the properties of the clusters with the projected track states at the front
face of the calorimeter. Linear and helix fits to the clusters and tracks are used to help make the correct associations.}
\item{If the energy of a calorimeter cluster does not agree with the associated track momentum, the cluster can be reconfigured by the statistical reclustering algorithms. 
The relevant calorimeter cells can be passed to a series of differently configured clustering algorithms to see if a configuration with better track-cluster compatibility can be found.}
\item{Fragment-removal algorithms look for neutral clusters (no track-association) that are actually fragments of nearby charged clusters (with track-associations). The
algorithms look for evidence of association between nearby neutral and charged clusters and evaluate the changes in track-cluster compatibility that would occur if the clusters were merged.}
\item{Particle flow objects (PFOs) are formed. If a particle contains tracks and associated clusters, the particle properties are extracted from the tracks. For neutral particles,
the calorimeter information is used.}
\item{Particle identification algorithms flag the reconstructed particles with PDG codes, identifying charged leptons. Photon identification is considered throughout the algorithms, but can be 
finalised at this stage.}
\end{itemize}

These algorithms have provided the particle flow reconstruction for the majority of studies performed for the ILD Detailed Baseline Design and the CLIC Conceptual Design 
Report~\cite{CLIC_PhysDet_CDR,CLICCDR_vol3}.

\section{Particle Flow Performance at ILC}
\label{performanceILC}
To assess the performance of the Pandora particle flow reconstruction algorithms, events reconstructed in a full GEANT4~\cite{Allison2006, Agostinelli2003} simulation of the detector model 
ILD\_o1\_v05~\cite{ILD} were used. For this purpose, $Z'$ particles were generated at energies of 91, 200, 360 and 500\,GeV. These are off-shell Z bosons, 
produced at rest at different centre-of-mass energies, which decay into light quarks and typically provide two back-to-back mono-energetic jets. The Pandora algorithms attempt to
reconstruct all the visible particles in each event. In order to avoid a bias from jet reconstruction, the sum of the reconstructed particle energy was analysed, $E_{jj}$. The 
resolution of the jet energy $E_{j}$ was then obtained by dividing $\mathrm{RMS}_{90}(E_{jj})$ by $\mathrm{mean}_{90}(E_{jj})$ and multiplying by $\sqrt{2}$:
\begin{equation}
   \frac{\mathrm{RMS}_{90}(E_j)}{\mathrm{mean}_{90}(E_{j})} = \frac{\mathrm{RMS}_{90}(E_{jj})}{\mathrm{mean}_{90}(E_{jj})} \sqrt{2}
\end{equation}

\begin{figure}[htb]
 \begin{center}
    \includegraphics[width=0.49\textwidth]{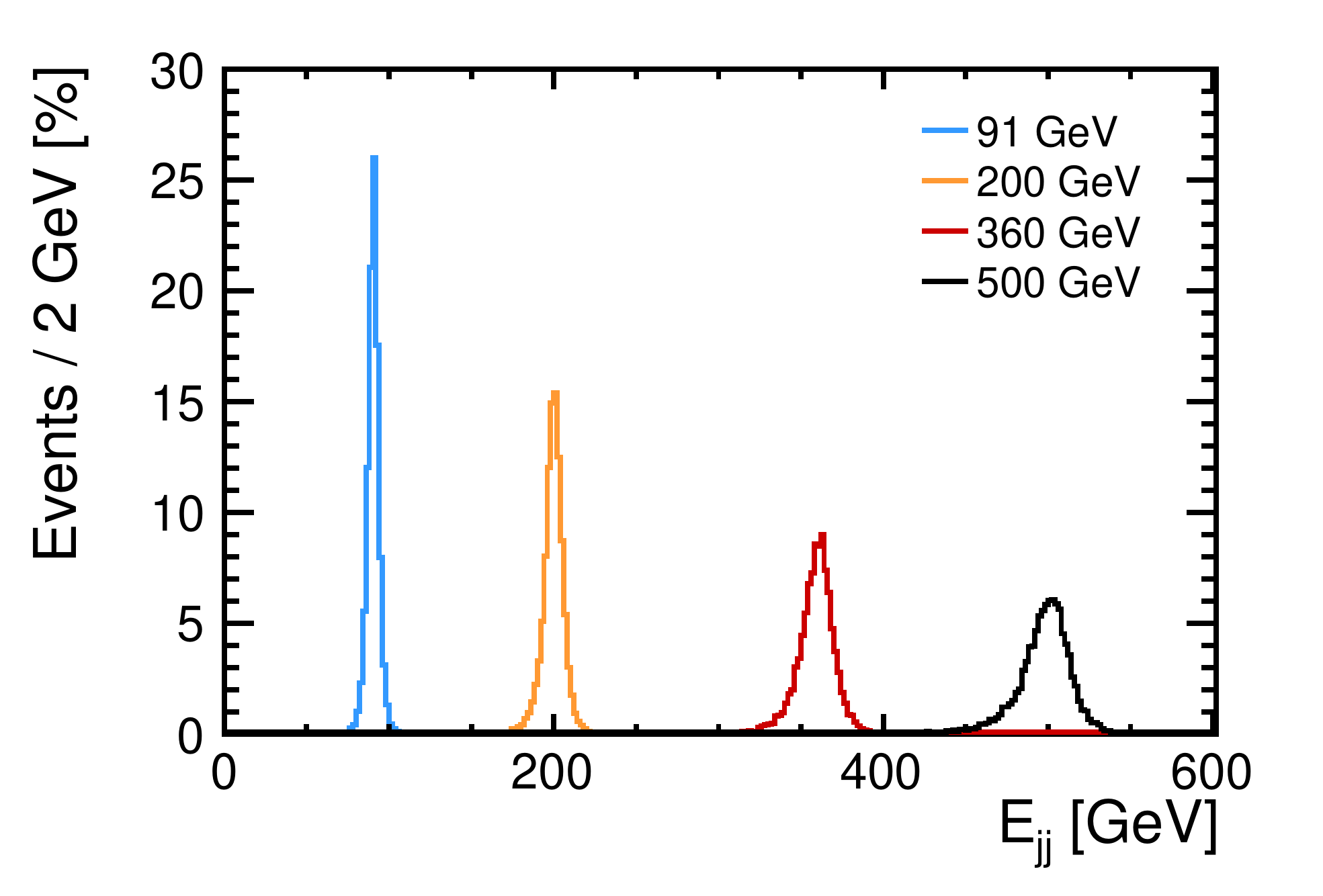}
    \includegraphics[width=0.49\textwidth]{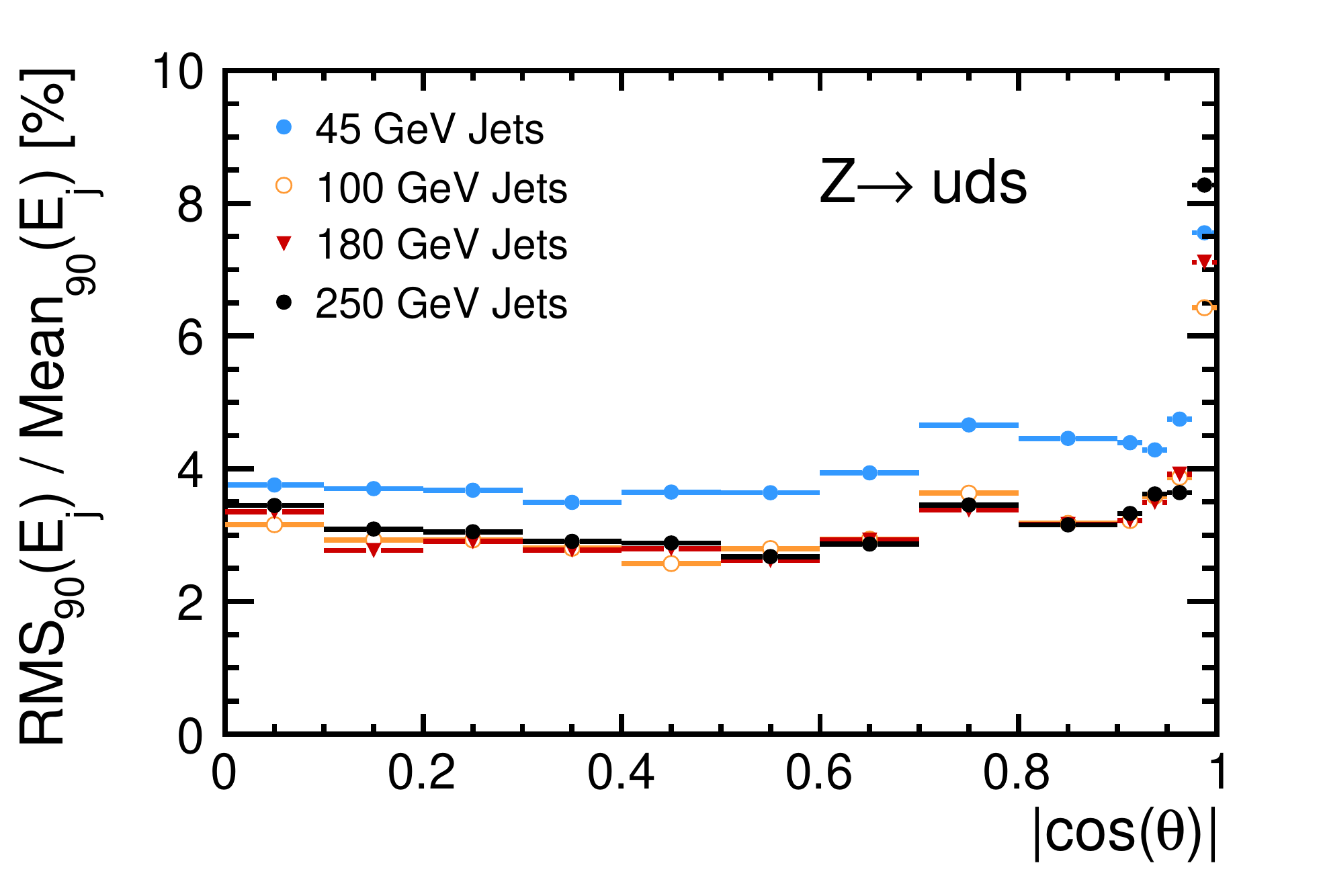}
   \caption{
         ({\it left}) The total reconstructed energy in $Z'$ events with different centre-of-mass energies in ILD\_o1\_v05,
         ({\it right}) The jet energy resolutions as a function of $|\cos(\theta)|$.
         }
   \label{fig5}
 \end{center}
\end{figure}

Figure~\ref{fig5} shows the total reconstructed energy for the $Z'$ particles and the jet energy resolutions as a function of the cosine of the polar angle. The challenging jet energy
resolution goal of \mbox{$\sigma_{E}/E\lesssim3.5\,\%$} is clearly surpassed for $100-250$\,GeV jets in all but the far-forward region of the detector. For 45\,GeV jets, the jet energy
resolution remains impressive, but is limited by the intrinsic calorimeter resolution to $\sim3.7\,\%$. For all energies, the resolutions remain effectively constant throughout the 
barrel region of the detector. The mean jet energy resolutions for the barrel region, $|cos(\theta)|<0.7$, are shown in Table~\ref{tab1}.

\begin{table}[!h]
\begin{center}
\begin{tabular}{ | r | c | c | c | c |}
\hline
\bf{Jet Energy, $\bf{E}_{j}$}  & \bf{45.6\,GeV} & \bf{100\,GeV} & \bf{180\,GeV} & \bf{250\,GeV} \\
\hline
$\bf{\mathrm{\bf{RMS}}_{90}(E_{j})/\mathrm{\bf{mean}}_{90}(E_{j})}$ & $3.66\pm0.05$ & $2.83\pm0.04$ & $2.86\pm0.04$ & $2.95\pm0.45$\\
\hline
\end{tabular}
\caption{Mean jet energy resolutions for the barrel region of ILD\_o1\_v05, $|cos(\theta)|<0.7$.}
\label{tab1} 
\end{center}
\end{table}

In order to better understand the different contributions to the jet energy resolution, it is possible to switch some of the standard Pandora algorithms with versions that use 
Monte Carlo (MC) information in order to `cheat' specific aspects of the reconstruction. One example would be a perfect photon reconstruction algorithm, which uses the links
between calorimeter cells and MC particles in order to identify all the true photon clusters and successfully separate them from other nearby particles. By running the Pandora
reconstruction with a series of MC algorithms, it is possible to assess the strengths and weaknesses of the standard reconstruction.

\begin{figure}[htb]
 \begin{center}
    \includegraphics[width=0.49\textwidth]{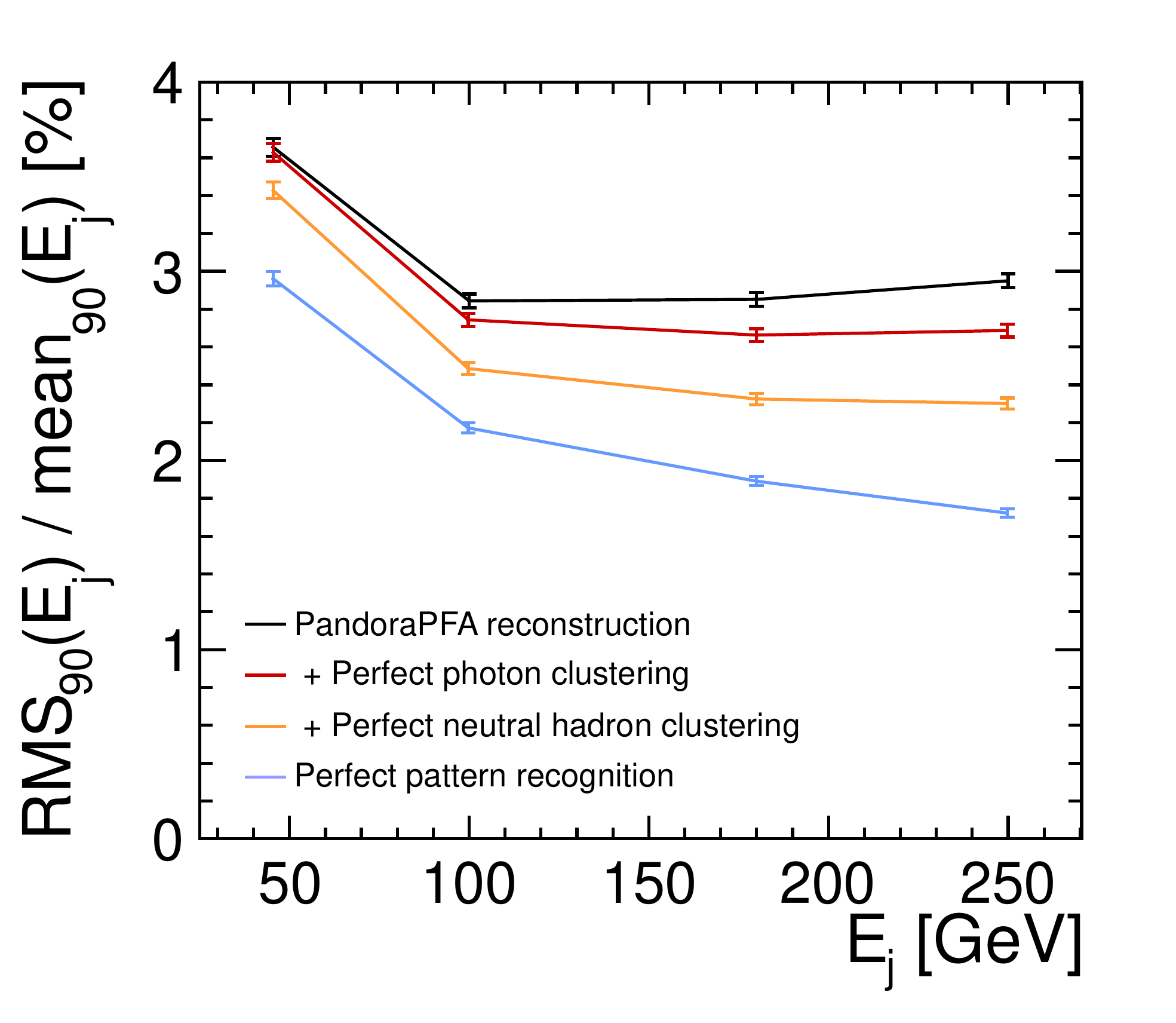}
    \includegraphics[width=0.49\textwidth]{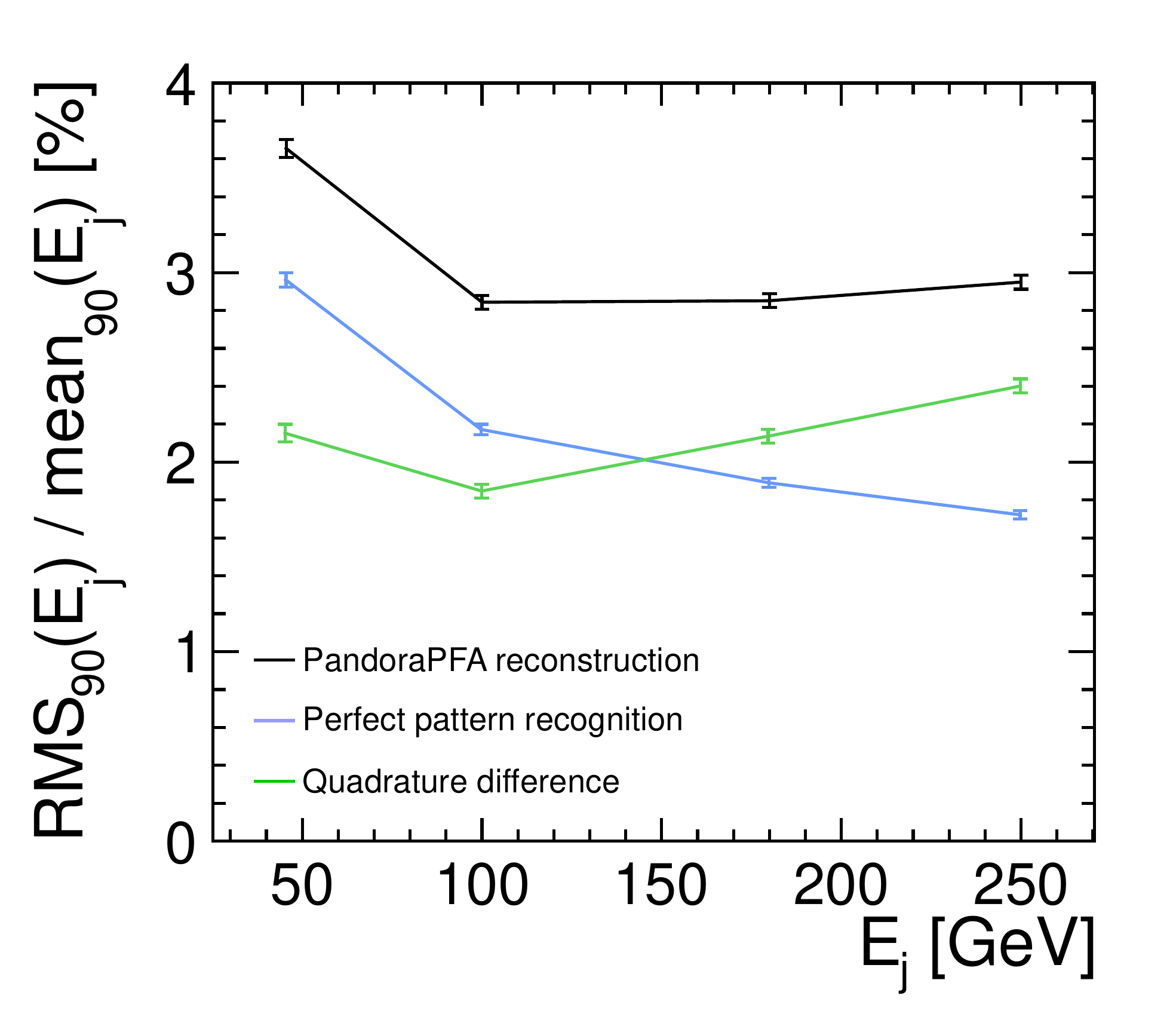}
   \caption{
         ({\it left}) The jet energy resolution as a function of jet energy for a series of different Pandora algorithm configurations.
         ({\it right}) The different contributions to the jet energy resolution as a function of the jet energy.
         }
   \label{fig6}
 \end{center}
\end{figure}

Figure \ref{fig6} shows the jet energy resolutions as a function of jet energy (considering only jets in the barrel region), for a series of different Pandora algorithm configurations.
Resolutions are shown for the standard reconstruction, with perfect photon clustering, with perfect photon and neutral hadron clustering and then with perfect pattern recognition.
For all the configurations, the reconstructed calorimeter energies and track momenta were still used to provide the four-vectors for the reconstructed particles. 

The performance with the perfect pattern recognition highlights the contribution to the resolution that is due to the intrinsic calorimeter (and inner tracker) resolutions. The
(quadrature) difference between the standard and perfect pattern recognition resolutions represents the contribution that is due to confusion in the events. It can be seen that
the main performance driver for the jet energy resolution varies with jet energy. For low energy jets, resolutions are limited by the intrinsic calorimeter resolutions. For high
energy jets, resolutions are limited by the confusion. The two contributions cross-over in the jet energy range $100-180$\,GeV.

Studies using five very different GEANT4 physics lists~\cite{Thomson:2009rp} suggest that the Particle Flow reconstruction is rather robust to the modelling of hadronic showers.

\section{Particle Flow Performance at CLIC}
The use of particle flow calorimetry at the proposed multi-TeV Compact Linear Collider poses a number of significant new challenges. At higher energies, detector occupancies
increase, and it becomes increasingly difficult to resolve energy deposits from individual particles. The experimental conditions at CLIC are also notably more challenging than
those at previous e$^{+}$e$^{-}$ colliders, with a bunch spacing of only 0.5\,ns and increased levels of beam-induced backgrounds. 

The modifications to the Pandora algorithms in order to improve the jet energy reconstruction for jet energies above 250\,GeV are documented in \cite{Marshall:2012ry}. This same 
reference also introduces a combination of timing and $\PT$ cuts that can be applied to reconstructed particles in order to significantly reduce the beam-induced background (a strategy 
that is only possible with the fine granularity particle flow approach). Finally, the reference presents the performance of Pandora particle flow calorimetry at CLIC and some 
of these results are reviewed below.

\begin{figure}[htb]
 \begin{center}
    \includegraphics[width=0.49\textwidth]{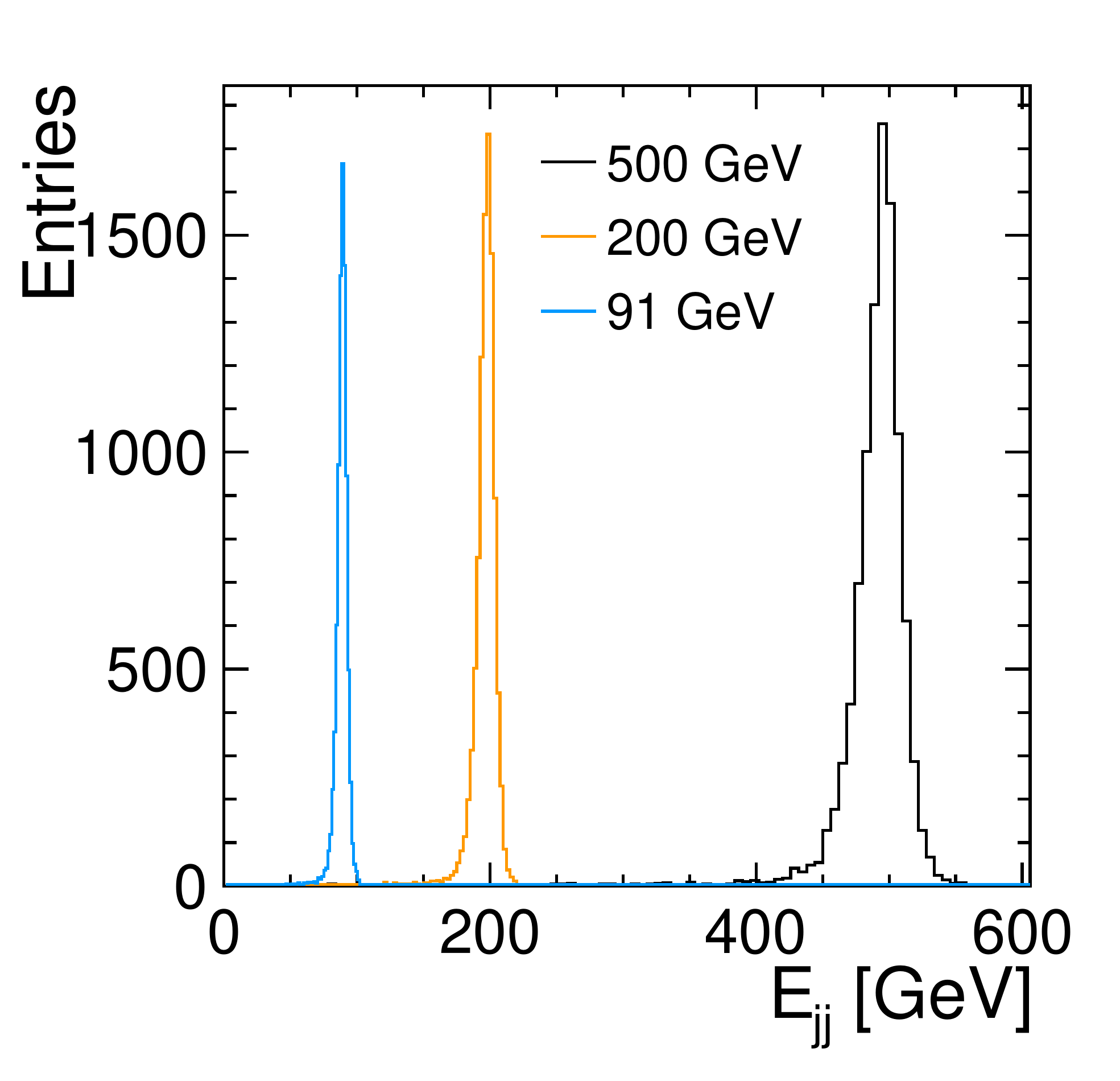}
    \includegraphics[width=0.49\textwidth]{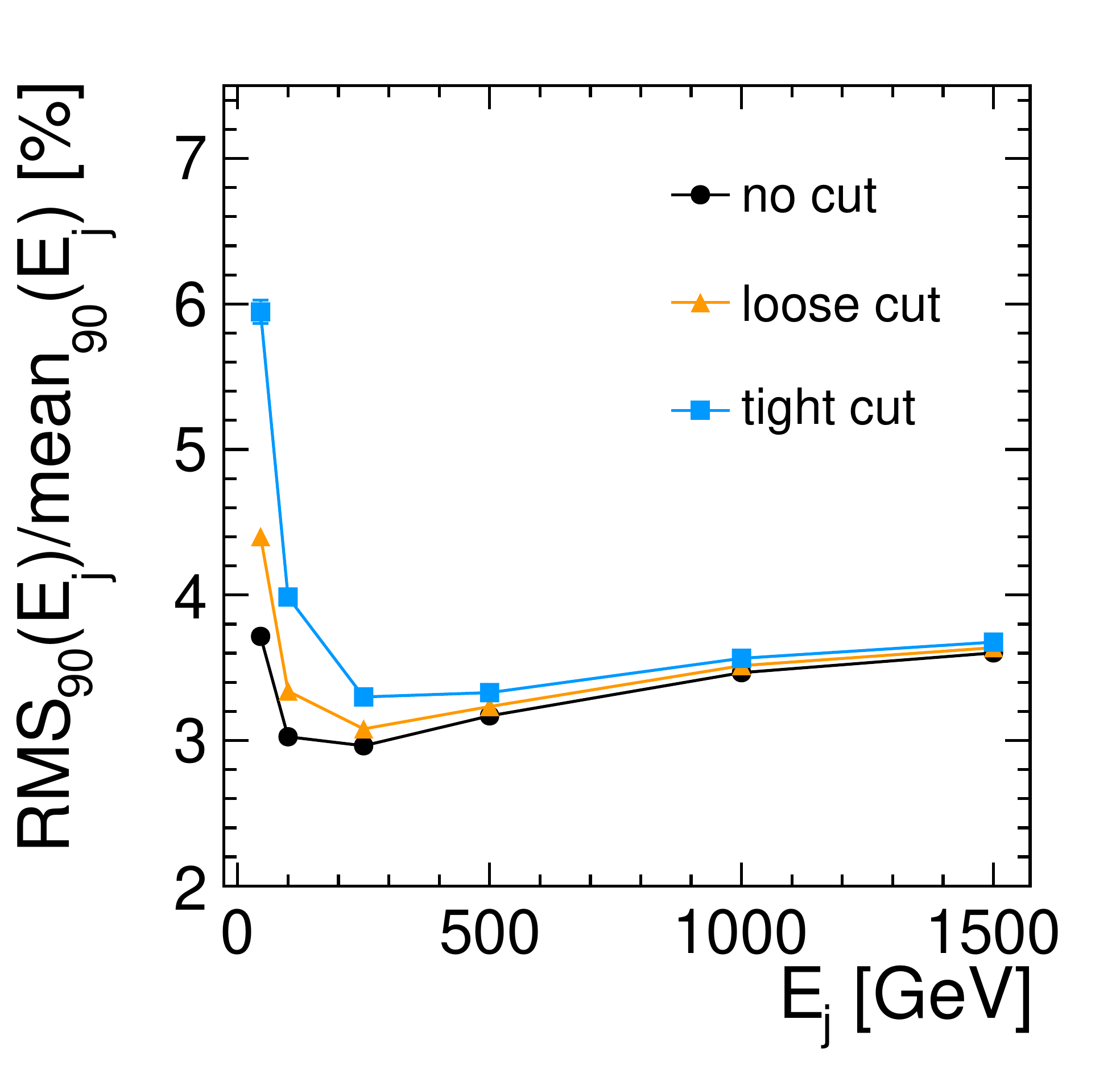}
   \caption{
         ({\it left}) The total reconstructed energy in $Z'$ events with different centre-of-mass energies in CLIC\_ILD\_CDR,
         ({\it right}) The jet energy resolutions as a function of jet energy with different background rejection cuts applied. Note that no backgrounds are actually present in the $Z'$ events.
          Source~\cite{Marshall:2012ry}.
         }
   \label{fig7}
 \end{center}
\end{figure}

Jet energy resolutions at CLIC are assessed using $Z'$ events, following the procedure described in Section~\ref{performanceILC}. Figure~\ref{fig7} shows the total reconstructed
event energy for three centre-of-mass energies. The Figure also shows the jet energy resolution as a function of the jet energy, for different configurations of background rejection cuts.
Without any background rejection cuts, the jet energy resolution is better than 3.7\,\% for jet energies in the wide range of 45\,GeV$-$1.5\,TeV. Note that no backgrounds are
present in the $Z'$ events, but the different background rejection cuts are applied in order to assess the impact of the cuts on underlying physics events. The cuts have a significant impact 
at low jet energies, but negligible effect at high jet energies.

In order to study the physics performance achievable using particle flow calorimetry at CLIC, the ability to distinguish between the hadronic decays of $W$ and $Z$ bosons was
examined. The $W$ events used for this study each contained two $W$ bosons, one of which decayed into a muon and neutrino, whilst the other decayed into quarks. Events were
generated for $W$ energies of 125-1000\,GeV, without background, with the nominal $\gamma\gamma\rightarrow$hadrons background and with twice the nominal background. The muon
was carefully removed and the $kt$ jet reconstruction algorithm~\cite{Fastjet} was used to force the events into two jets. When background was present, the tight background 
rejection cuts were applied. The di-jet energy and mass resolutions were examined and were found to be comparable to the resolutions observed for $Z'$ events.

\begin{figure}[htb]
 \begin{center}
    \includegraphics[width=0.49\textwidth]{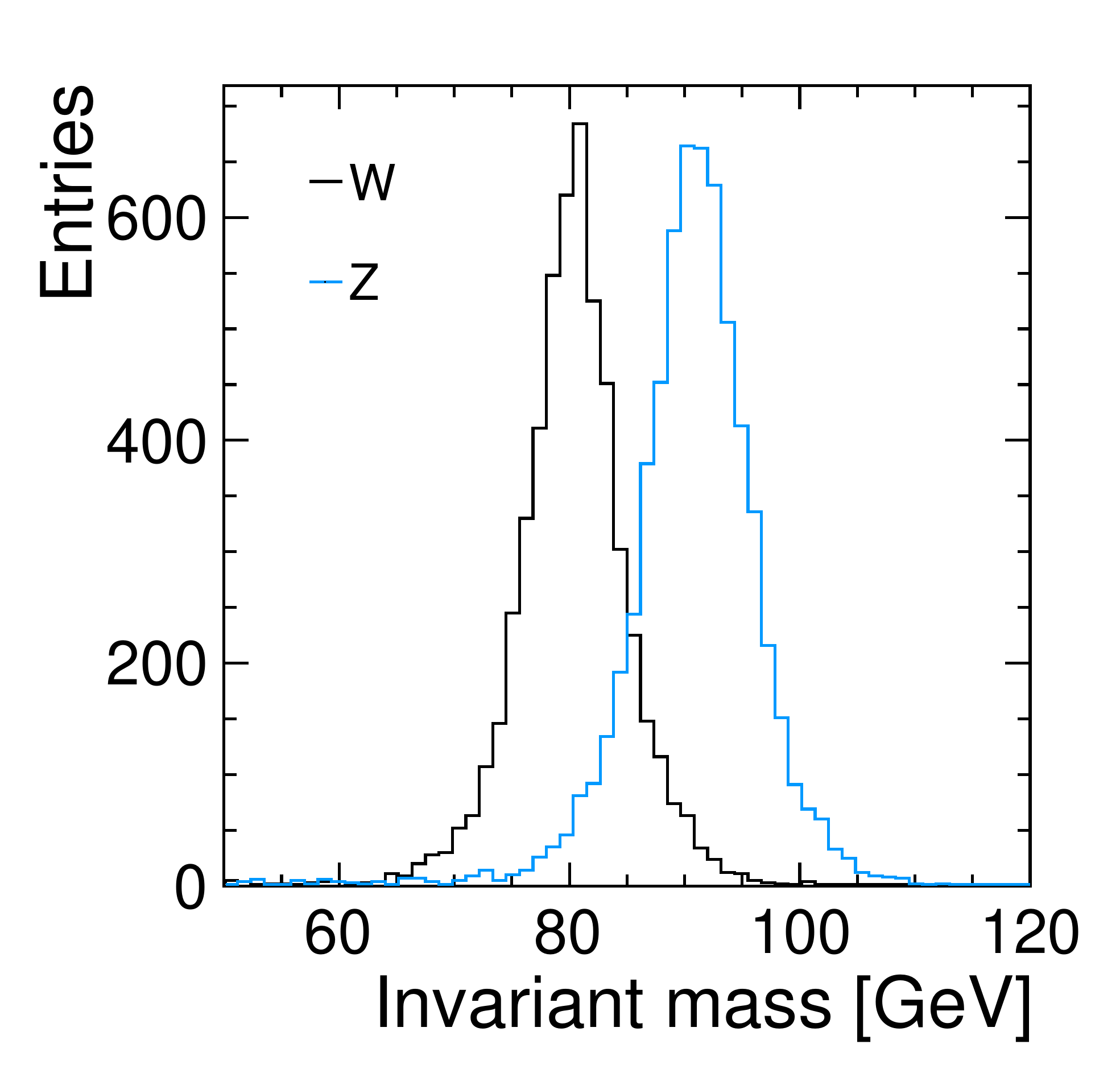}
    \includegraphics[width=0.49\textwidth]{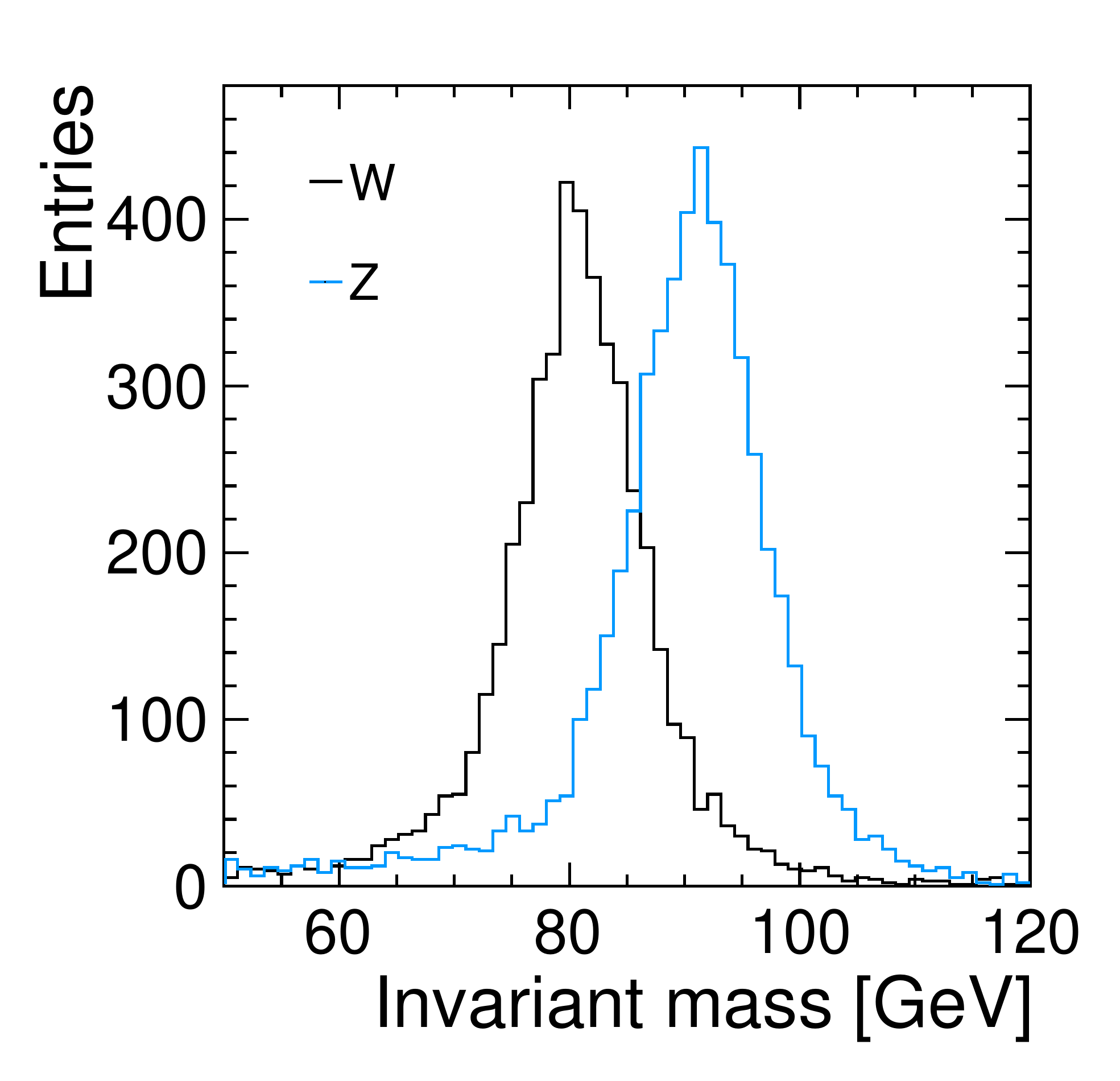}
   \caption{
         Mass distributions for reconstructed $W$ and $Z$ at an energy of 500\,GeV, without background ({\it left}), and with the nominal background ({\it right}). Source~\cite{Marshall:2012ry}.
         }
   \label{fig8}
 \end{center}
\end{figure}

The $Z$ events used for the study each contained two $Z$ bosons, one of which decayed to neutrinos, whilst the other decayed into quarks. Events were available with the same energies
and background configurations as for the $W$ sample. The $W$ and $Z$ samples were treated in exactly the same way (with the minor exception of muon removal for the $W$ events)
and the resulting di-jet invariant masses were compared. Figure~\ref{fig8} shows the mass distributions of the reconstructed $W$ and $Z$ at an energy of 500\,GeV, without background
and with the nominal $\gamma\gamma\rightarrow$hadrons background. Without background, there is clear separation between the peaks. With background, the separation is slightly
degraded.

\begin{figure}[h!] 
\begin{center}
\includegraphics[width=.49\textwidth]{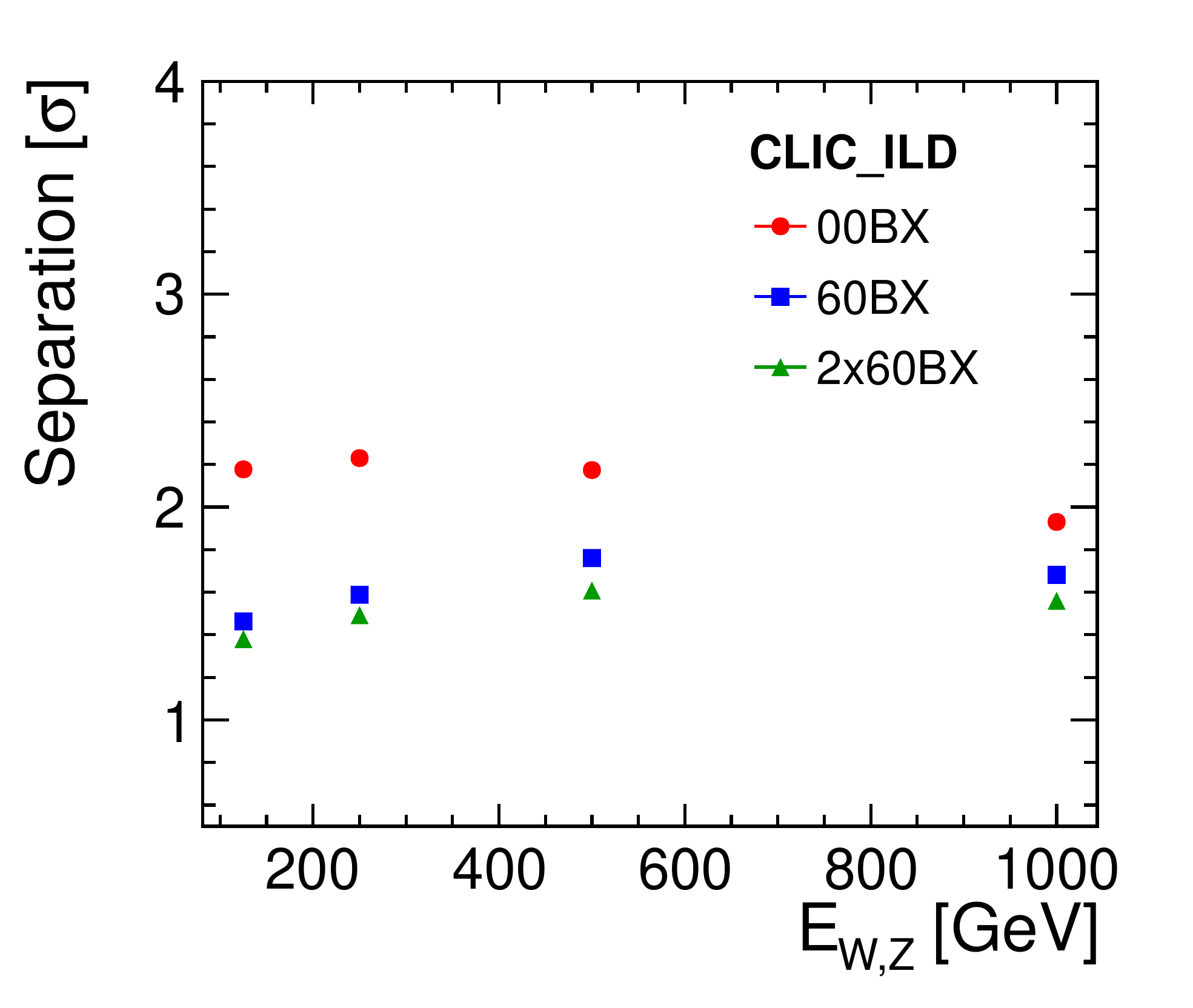}
\caption{The separation of the $W$ and $Z$ peaks achieved as a function of the gauge boson energy, for three different background configurations.} 
\label{fig9} 
\end{center}
\end{figure}

The separation of the $W$ and $Z$ peaks was quantified by determining the fraction of mis-identified events for the optimum di-jet invariant mass cut. The natural widths of the
$W$ and $Z$ boson restrict the efficiency to $<94\,\%$~\cite{Thomson:2009rp}. The fraction of mis-identified events is converted into an equivalent Gaussian separation. Figure~\ref{fig9}
shows the separation as a function of the gauge boson energy for the different background configurations. Without background, a $2\,\sigma$ separation is maintained across the energy
range 125\,GeV$-$1\,TeV. With nominal background, this falls to $1.7\,\sigma$.

It would be exceedingly difficult to achieve this level of $W$ and $Z$ separation using a traditional approach to calorimetry, even without background. When background is included,
particle flow calorimetry proves to be invaluable, with the reconstruction of individual particles allowing classification of the particles as being from background or from the 
underlying interaction.


\bibliography{mybib}{}
\bibliographystyle{unsrt}

\end{document}